\documentclass{sf2a-conf2021}
\usepackage{graphicx}
\usepackage{hyperref}
\usepackage[]{natbib}  
\usepackage{epstopdf}
\usepackage{bm}
\usepackage{amsmath}
\usepackage{amssymb}

\def\BibTeX{{\rm B\kern-.05em{\sc i\kern-.025em b}\kern-.08em
    T\kern-.1667em\lower.7ex\hbox{E}\kern-.125emX}}
\bibpunct{(}{)}{;}{a}{}{,}  %%%%%%%%%%%%%  A&A bibliography style
\newcommand{\mI}[1]{\mathcal{I}_\mathrm{#1}}
\newcommand{\uw}{\bm u_\mathrm{w}}
\newcommand{\unw}{\bm u_\mathrm{nw}}
\newcommand{\bn}{\bm\nabla}
\newcommand{\Ek}{\mathrm{Ek}}
\newcommand{\ft}{\bm f_\mathrm{t}}
\newcommand{\e}[1]{\mathrm{e}^{#1}}
\begin{document}

\TitreGlobal{SF2A 2021}

%%-----------------------------------------------------------------
%%      the top matter
%%

\title{Nonlinear simulations of tides in the convective envelopes \\of low-mass stars and giant gaseous planets}

\runningtitle{Nonlinear simulations of inertial waves in a convective shell}

\author{A. Astoul}\address{Department of Applied Mathematics, School of Mathematics, University of Leeds, Leeds, LS2 9JT, UK}

\author{A. J. Barker$^1$}

%% IF Author3 has the same affiliation than Author1:
%\author{C.\,E. Author3$^1$}

%% IF Author3 has its own affiliation:
%\author{C.\,E. Author3}\address{Dept. of Chess, University of Games, 35101 Las Vegas, Monaco} 

%% Keep this line, even if the page will be settled afterwards.
\setcounter{page}{237}

%%-----------------------------------------------------------------

\maketitle

%%-----------------------------------------------------------------
%%        The abstract
%% 
%%  Warning!  within the abstract:
%%  - do not use macros. 
%%  - do not use commands like: \cite, \citet, \citep ... etc.

\begin{abstract}
In close two-body astrophysical systems, such as binary stars or Hot Jupiter systems, tidal interactions often drive
dynamical evolution on secular timescales. Many host stars and presumably giant gaseous planets feature a convective envelope. Tidal flows 
generated therein by the tidal potential of the companion 
can be dissipated through viscous friction, leading to the redistribution and exchange of angular momentum within the convective shell and with the companion, respectively. In the tightest systems, nonlinear effects are likely to have a significant impact on the tidal dissipation and trigger differential rotation in the form of zonal flows, as has been shown in previous studies. In this context, we investigate how the addition of nonlinearities affect the tidal flow properties, and energy and angular momentum balances, using 3D nonlinear simulations of an adiabatic and incompressible convective shell. In our study, we have chosen a body forcing where the equilibrium tide (the quasi-hydrostatic tidal flow component) acts via an effective forcing to excite tidal inertial waves in a spherical shell. %, while using stress-free boundary conditions.  %Too much detail for abstract?
Within this set-up, we show new results for the amplitude of the energy stored in zonal flows, angular momentum evolution, and its consequences on tidal dissipation in the envelopes of low-mass stars and giant gaseous planets.
\end{abstract}

%% Insert the keywords (to appear in the ADS indexing)
%% Keywords must be separated by a comma

%%-----------------------------------------------------------------

\section{Introduction}
%%---------------------
In tight stellar and exoplanetary systems, tidal interactions are a key process to understand orbital migration %decay -- can be outwards? ha yes
and circularisation, spin synchronisation, and possibly the low obliquities of the closest bodies \citep[e.g.][]{M2019,O2014}. %,O2020}. remove? No strong opinion! ok
For low-mass stars and Hot Jupiter planets which feature a convective envelope, the dissipation of inertial waves restored by the Coriolis acceleration significantly contributes to the tidal dissipation. This is particularly true in the early stages of the life of rapidly rotating low-mass stars, as shown for example in \cite{M2015}, \cite{BM2016}, %\cite{GB2017}
and \cite{B2020}, and possibly may explain the rapid orbital expansion measured for Titan through resonant-locking of inertial modes in Jupiter \citep{LC2020}. Furthermore, for the most compact two-body systems, internal nonlinear fluid effects can become important \citep{BO2016}, especially when considering the small-scales of inertial waves in spherical shells %as inertial attractor modes : ok 
\citep{RG2001}.
This typically happens when the tidal amplitude parameter\footnote{It is a measure of the ratio of the tidal gravity to the self-gravity at the surface of the perturbed body \citep[e.g.][]{BO2016}.} $\epsilon$ is large, as for the ultra short-period Hot Jupiter WASP-19 b \citep[$\epsilon\sim0.05$, e.g.][]{O2014}.
 In \citet[denoted Paper~I in the following]{FB2014}, the authors performed a nonlinear numerical study of tidally-forced inertial waves in spherical shells. They found that the amplitude of the tidal dissipation could greatly differ from  linear tidal response predictions \citep[e.g. as in][]{O2009} due to the inhomogeneous deposition of angular momentum which generates strong azimuthal zonal flows \citep[see also][]{A2021}. They also observed for some forcing frequencies unexpected evolution of the angular momentum, pushing the body away from synchronisation. Building upon their initial study, we have performed  nonlinear hydrodynamical simulations of tidal waves in an adiabatic and incompressible convective envelope, but with different boundary conditions and tidal forcing. While they imposed a radial velocity through the outer boundary to force the tidal flow, we use an effective body force which  accounts for the residual action of 
%mimics the presence
the large-scale non-wavelike tidal flow in exciting inertial waves \citep{O2013}, %\citep[][]{O2005,O2013},,
with stress-free impenetrable boundary conditions at the inner and outer boundaries. Within this more realistic set-up, we are able to separate wavelike and non-wavelike contributions in the nonlinear terms to elucidate the unphysical behaviour obtained for certain frequencies in Paper I. %highlighting a flux term in the energy balance leading to a wrong evolution of the total angular momentum. 
By removing the relevant nonlinearities that produce this unphysical behaviour (which should be less important than the ones we retain in reality), %flux term, 
we present new results regarding the amplitude and spatial structure of kinetic energy, tidal dissipation, and azimuthal component of the velocity of inertial waves propagating in an adiabatic and incompressible convective shell.

%\cite{O2014,B2016}

\begin{figure}[t!]
 \centering
 \includegraphics[width=0.329\textwidth,clip]{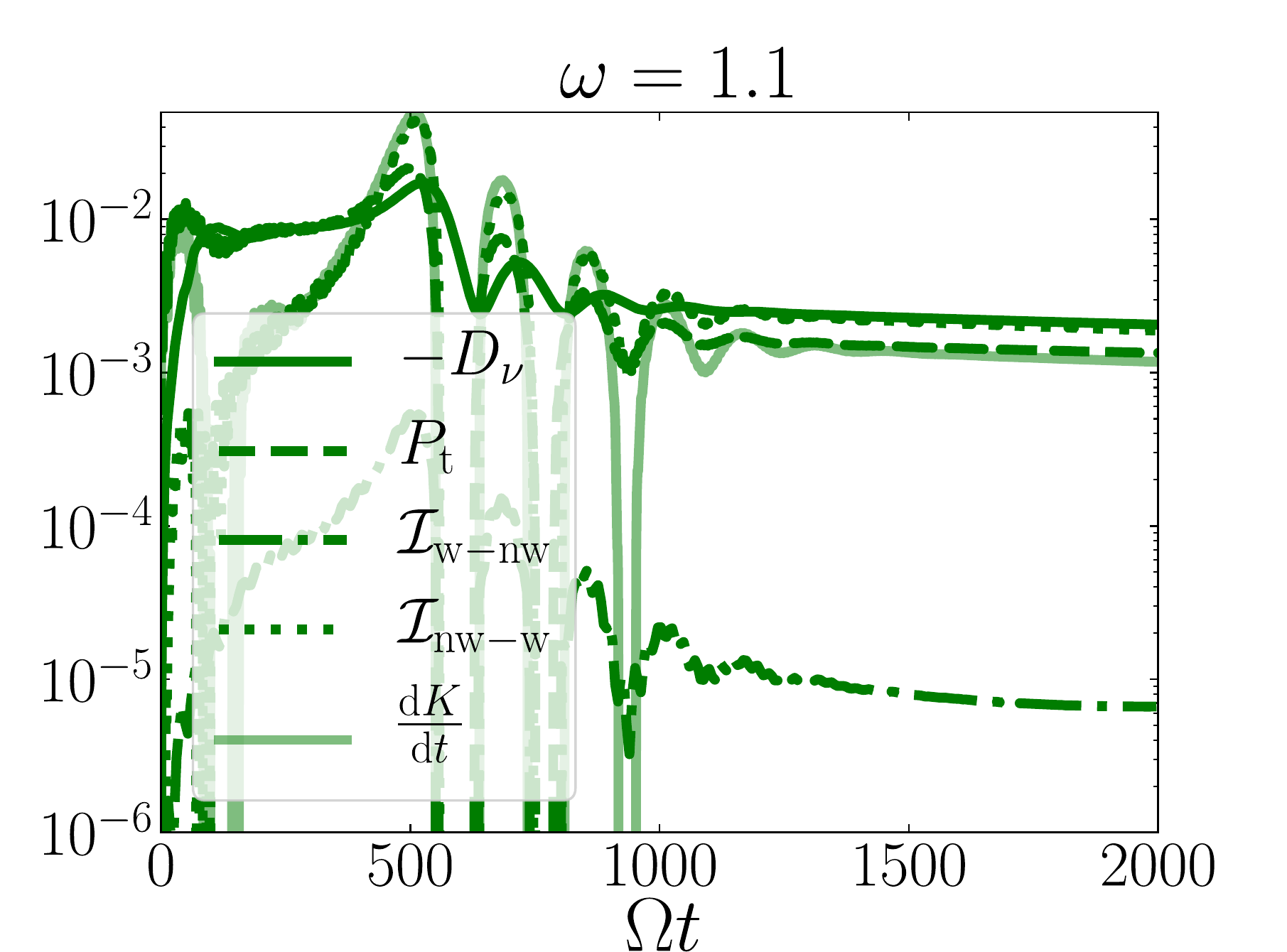}      
 \includegraphics[width=0.329\textwidth,clip]{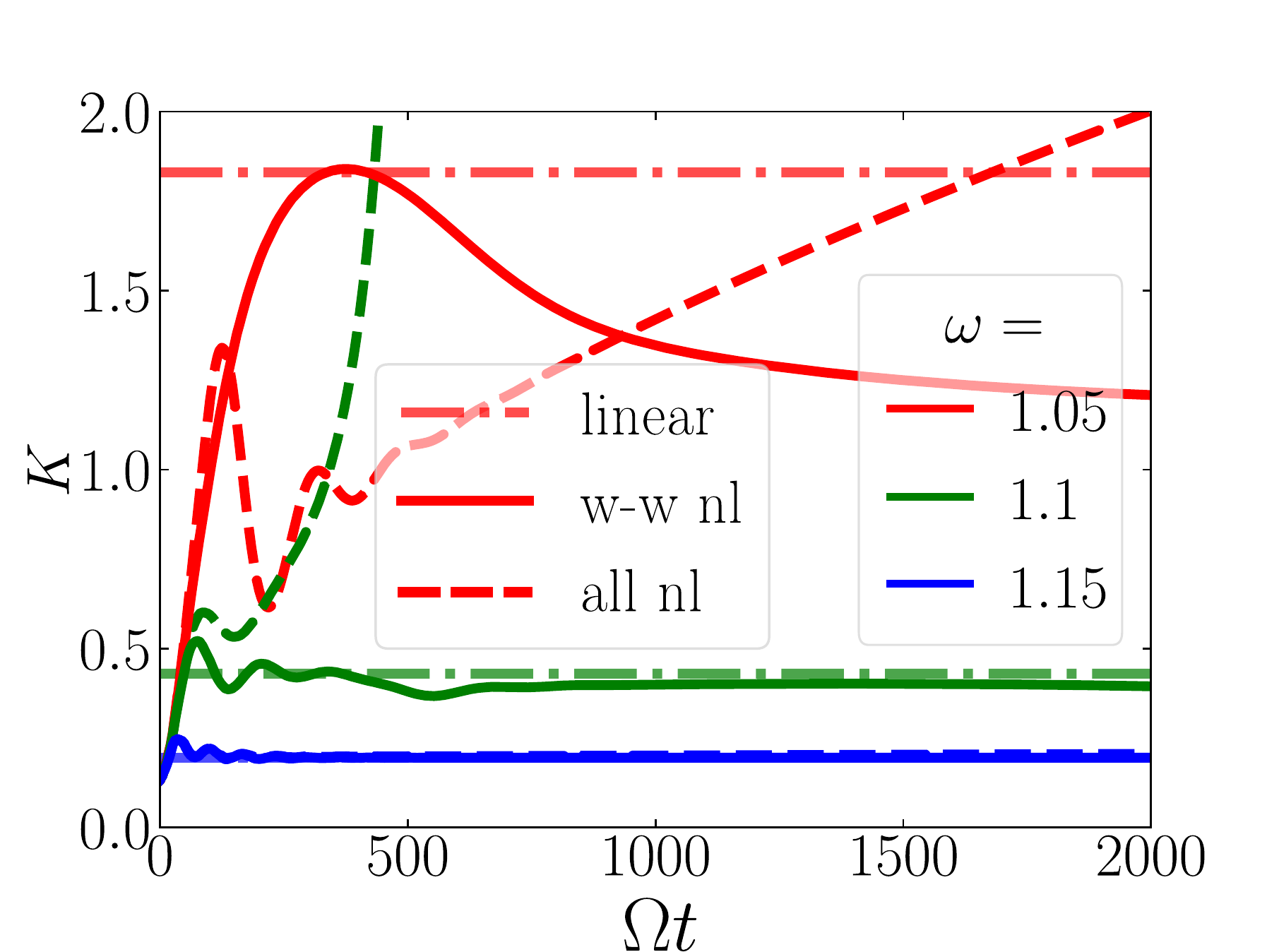}      
 \includegraphics[width=0.329\textwidth,clip]{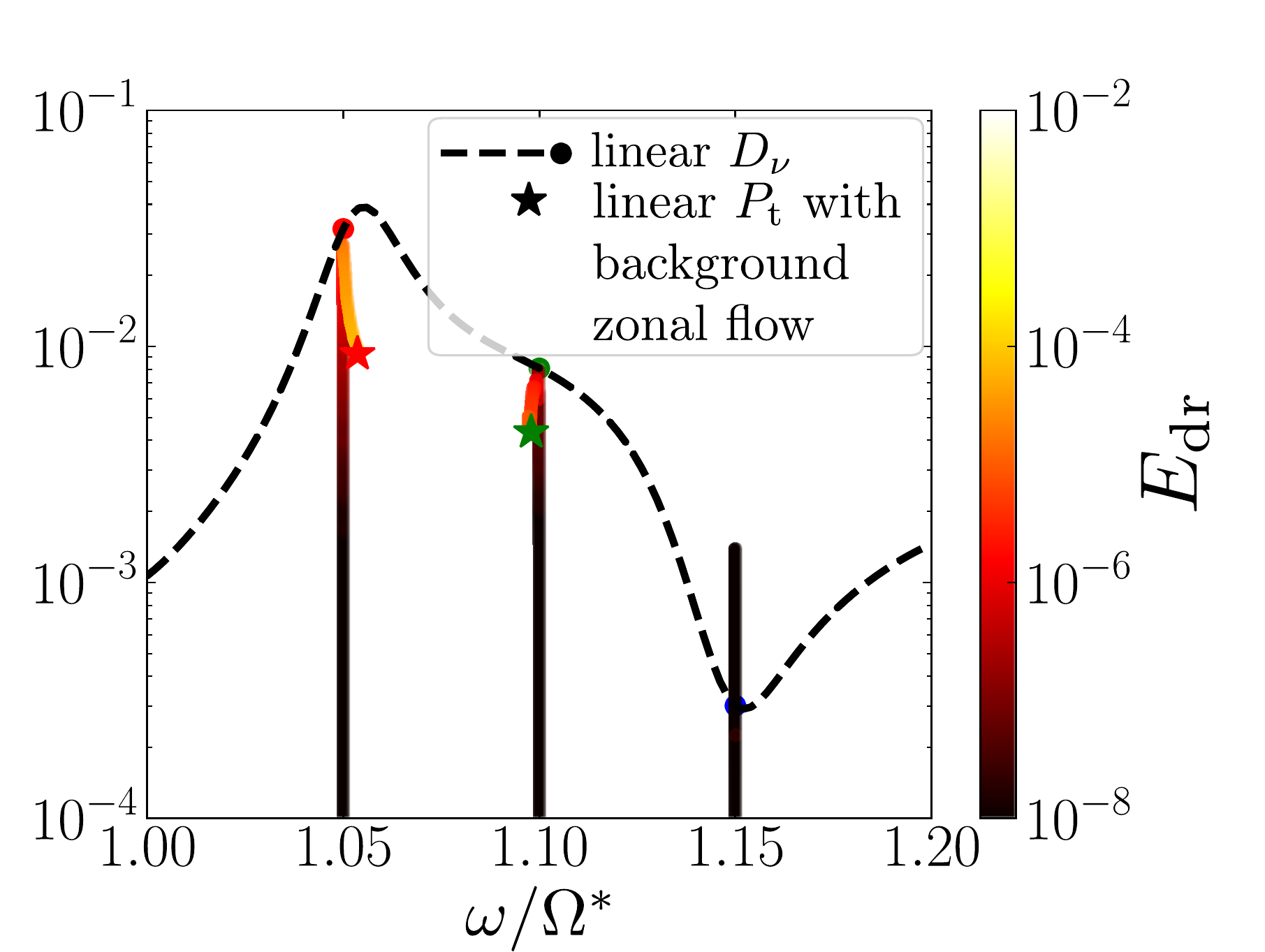}      
 \caption{{\bf Left:} Terms in the energy balance (Eq. (\ref{eq:rey})) over time.  {\bf Middle:} Kinetic energy over time
 %for three different initial forcing frequencies
 with or without the various nonlinear terms (nl).  {\bf Right:} Dissipation spectrum $D_\nu$ of wavelike-wavelike nonlinear simulations as a function of tidal forcing frequency in the fluid frame $\omega/\Omega^*$, where $\Omega^*$ is the modified spin rate due to the creation of zonal flows (see Paper~1 for a definition), also showing the energy in the differential rotation $\mathrm{E_{dr}}$. In the three panels, $C\approx0.009$.
 %The star symbols indicate the results of new linear simulations taking the zonal flow as a background flow. 
% Both panels have $C\approx0.009$.
  }
  \label{astoul:fig1}
\end{figure}
\section{Governing equations and energy balance for nonlinear tidal inertial waves}
We model the outer convective envelope of our low-mass star or giant gaseous planet as an incompressible and adiabatic spherical shell as a first approach to treat the nonlinear hydrodynamic tidal response. The action of turbulent convection on tides is considered through a viscous term $\nu\Delta\bm u$  \citep[e.g.][and references therein]{Duguid2020b}, where $\nu$ is the (constant) effective/turbulent viscosity and $\bm u$ is the total %perturbed 
tidal flow. % after linearisation around a stationary background. 
The flow $\bm u$ is decomposed into a large-scale non-wavelike component $\unw$ (the equilibrium tide in this model) % incompressible assumption) 
and a wave-like component $\uw$ (the dynamical tide, consisting of inertial waves) %which are inertial waves in a rotating and adiabatic medium 
such that $\bm u=\unw+\uw$ as in \citet[building upon earlier ideas reviewed in e.g. \citealt{Z2008}]{O2013}. %\citep[building upon earlier ideas by e.g.][]{Z1966a,Z1966b,Z1966c,Z1970}. %Associated with stress-free and impenetrable boundary conditions, the $2^\mathrm{nd}$ order equation of motion and continuity for tidally forced inertial waves in the fluid frame is:
The equations of motion and continuity for tidally-forced inertial waves in the fluid frame are then:
\begin{align}
   \partial_t\uw+(\bm u\cdot\bn)\bm u+2\bm\Omega\wedge\uw
   &=-\frac{\bn p_\mathrm{w}}{\rho}+\ft+\nu\Delta\bm u, \label{eq:mo}\\
   \bn\cdot\uw& =0, \label{eq:con}
\end{align}
which are solved together with stress-free and impenetrable boundary conditions on the inner and outer boundaries at radii $r=\alpha R$ and $r=R$, respectively, with $\alpha$ the radial aspect ratio and $R$ the radius of the body.
We have introduced $\bm\Omega$, the initial uniform %mean 
rotation rate along the vertical axis, $p_\mathrm{w}$, the reduced wavelike pressure, %where the centrifugal acceleration can be included %omit?, ok, reduced pressure in the term used in Michel's book
and $\rho$, the (uniform) mean fluid density. The effective tidal forcing is $\ft = -2\bm\Omega\wedge\bm u_\mathrm{nw}$, which takes into account the remaining action of non-inertial terms (here only the Coriolis acceleration) on the non-wavelike flow, since the equilibrium tide satisfies a quasi-hydrostatic equation of motion. The non-wavelike flow can be written 
$\unw=\mathrm{Re}[C~i\omega\bn\left[f(r)Y^2_2(\theta,\varphi)\right]\e{-i\omega t}]$, where $C$ is proportional to  the tidal amplitude parameter $\epsilon$ (with a factor depending on the internal structure), $\omega$ is the tidal forcing frequency, $f(r)$ is a dimensionless radial function required for the scalar potential to satisfy Laplace's equation, and $Y^2_2(\theta,\varphi)$ is the dominant quadrupolar spherical harmonic function for colatitude $\theta$ and azimuth $\varphi$ \citep{O2013}. Our approach considers the non-wavelike flow to be an imposed ``background flow" to study the instantaneous tidal response (relative to tidal evolutionary timescales), as described in \cite{BA2021}. We use units of length and time equal to $R$ and $\Omega^{-1}$, respectively.

Taking the scalar product of Eq. (\ref{eq:mo}) with $\rho\uw$ and integrating spatially, we obtain the energy balance for tidal inertial waves:
\begin{equation}
    \partial_t{\langle K_\mathrm{w}\rangle_V}= %\mI{w-w}+\mI{nw-nw}+
    \mI{nw-w}+ \mI{w-nw}-D_\nu+P_\mathrm{t},
    \label{eq:rey}
\end{equation}
where $\langle K_\mathrm{w}\rangle_V=\langle \rho|{\uw}|^2/2\rangle_V$ is the wavelike kinetic energy integrated over the volume $V$,  $\mI{i-j}=-\langle\rho\uw\cdot(\bm u_i\cdot\bn)\bm u_j\rangle_V$, with $i,j\in\{\mathrm{w,nw}\}$, are the terms coming from the mixed (w-nw and nw-w) nonlinearities which transfer energy between the non-wavelike and wavelike flows, $D_\nu$ is the viscous %turbulent 
dissipation of wavelike flows\footnote{The dissipation of the non-wavelike flow has been found to be negligible in our simulations since $\nu$ is small, as is also typically expected in reality.} and the energy injected into tidal waves by the forcing is $P_\mathrm{t}=\langle\uw\cdot\ft\rangle_V$. Note that $\mI{i-i} =-\oint_{\partial V} K_\mathrm{i}\,\uw\cdot{\bm n}\mathrm{d} S=0$ with $\bm n$ the unit normal vector to the bounding surface $\partial V$, using the divergence theorem along with the incompressible and impenetrable assumptions (which can also be applied to demonstrate that $\langle \uw \cdot \nabla p_w\rangle_V=0$). One can also demonstrate that $\mI{nw-w}=-\oint_{\partial V} K_\mathrm{w}~\unw\cdot\bm n\mathrm{d} S$, which does not necessarily vanish in our idealised spherical model. However, in a more realistic ellipsoidal model, the streamlines of the non-wavelike tidal flow in the bulge frame are expected to be tangential to the boundary so that $\unw\cdot\bm n=0$, cancelling $\mI{nw-w}$ in that frame \citep[see also][]{BA2021}. The only physical nonlinear term expected to contribute to the energy exchange between inertial waves and the non-wavelike tidal flow is $\mI{w-nw}= -\langle \rho\uw\cdot(\uw\cdot\bn)\unw\rangle_V$, namely Reynolds stresses involving correlations between wavelike flow components.

%%---------------------------------------------------------------
\section{Analysis of nonlinear simulations}
%\begin{figure}[t!]
% \centering
% \includegraphics[width=0.49\textwidth,clip]{astoul_S6_fig3}      
% \includegraphics[width=0.49\textwidth,clip]{astoul_S6_fig4}      
% \caption{{\bf Left:} Kinetic energy over time for three different initial forcing frequencies (in colours), with or without the various nonlinear terms (shown using different line styles). {\bf Right:} Dissipation spectrum $D_\nu$ of wavelike-wavelike nonlinear simulations as a function of tidal forcing frequency in the fluid frame $\omega/\Omega^*$, where $\Omega^*$ is the modified spin rate due to the creation of zonal flows (see Paper~1 for a definition), also showing the energy in the differential rotation $\mathrm{E_{dr}}$ (colour). The star symbols indicate the results of new linear simulations taking the zonal flow as a background flow. Both panels have $C\approx0.009$.}
%  \label{astoul:fig2}
%\end{figure}

To solve Eqs. (\ref{eq:mo}) and (\ref{eq:con}), we use the pseudo-spectral code MagIC\footnote{\url{https://magic-sph.github.io/}} \citep[e.g.][]{CA2001} using Chebyshev polynomials in the radial direction and a spherical harmonic decomposition in the azimuthal and latitudinal directions, utilising the SHTns library for fast spherical harmonic transforms \citep{S2013}.
In these initial simulations, %the entropy is fixed at inner and outer boundary, %it is fixed everywhere! ha yes it's true since adiabatic~isentropic
the Ekman number is set to $\Ek=\nu/(\Omega R^2)=10^{-5}$, %FACTOROFTWONOTIN2014?, no indeed, bad habit %(with $R$ the radius of the body), 
the aspect ratio is $\alpha=0.5$, and we vary the tidal forcing frequency $\omega$ and the tidal amplitude $C$. Our typical spatial resolution is $N_r= 97$ Chebyshev points in radius and spherical harmonics up to degree $l_\mathrm{max} = 85$. % NPHI=256FORALL?. 
Further details about the simulations will be presented elsewhere.

In the left panel of Fig.~\ref{astoul:fig1}, we display each term in the energy balance (Eq. (\ref{eq:rey})) for $C\approx0.009$ and $\omega=1.1$. The term $\mI{nw-w}$ represents an unrealistic flux through the boundary which is not negligible compared to the tidal dissipation and tidal power $P_\mathrm{t}$, unlike the ``physical" transfer term $\mI{w-nw}$, which is very small here. This is especially true for low to moderate tidal amplitudes satisfying $C\lesssim0.05$,
%, as shown by the green and blue curves in the right panel(also found for other frequencies, not presented), 
but not for higher tidal amplitude %(black curves)
(e.g. $C\sim0.1$)
where $\mI{w-nw}$ can contribute more significantly. As a result, we have chosen to switch off nonlinear terms involving non-wavelike tides in the momentum equation, which is justified except for the highest tidal amplitudes considered ($C\lesssim0.05$). % since we also remove the term $\mI{w-nw}$.
Our choice is also motivated by the fact that the total angular momentum $\bm L=\langle\rho \bm r \wedge \bm u\rangle_V$ can be shown to
%$\bm L=\langle\bm r \wedge \bm u\rangle_V$
evolve as:
\begin{equation}
\partial_t\bm L=-%\frac{1}{V}
\oint_{\partial V}\rho(\bm r\wedge\bm u)\unw\cdot\bm n~\mathrm{d}S.
\end{equation}
Thus, $\bm L$ is not conserved in this model (as $\unw$ is perfectly maintained) only because of the unphysical non-wavelike flux through the spherical boundaries, which drives the unexpected evolution of the total angular momentum as in Fig. 17 of Paper~I. 

\begin{figure}[t!]
 \centering
 \includegraphics[width=0.24\textwidth,clip]{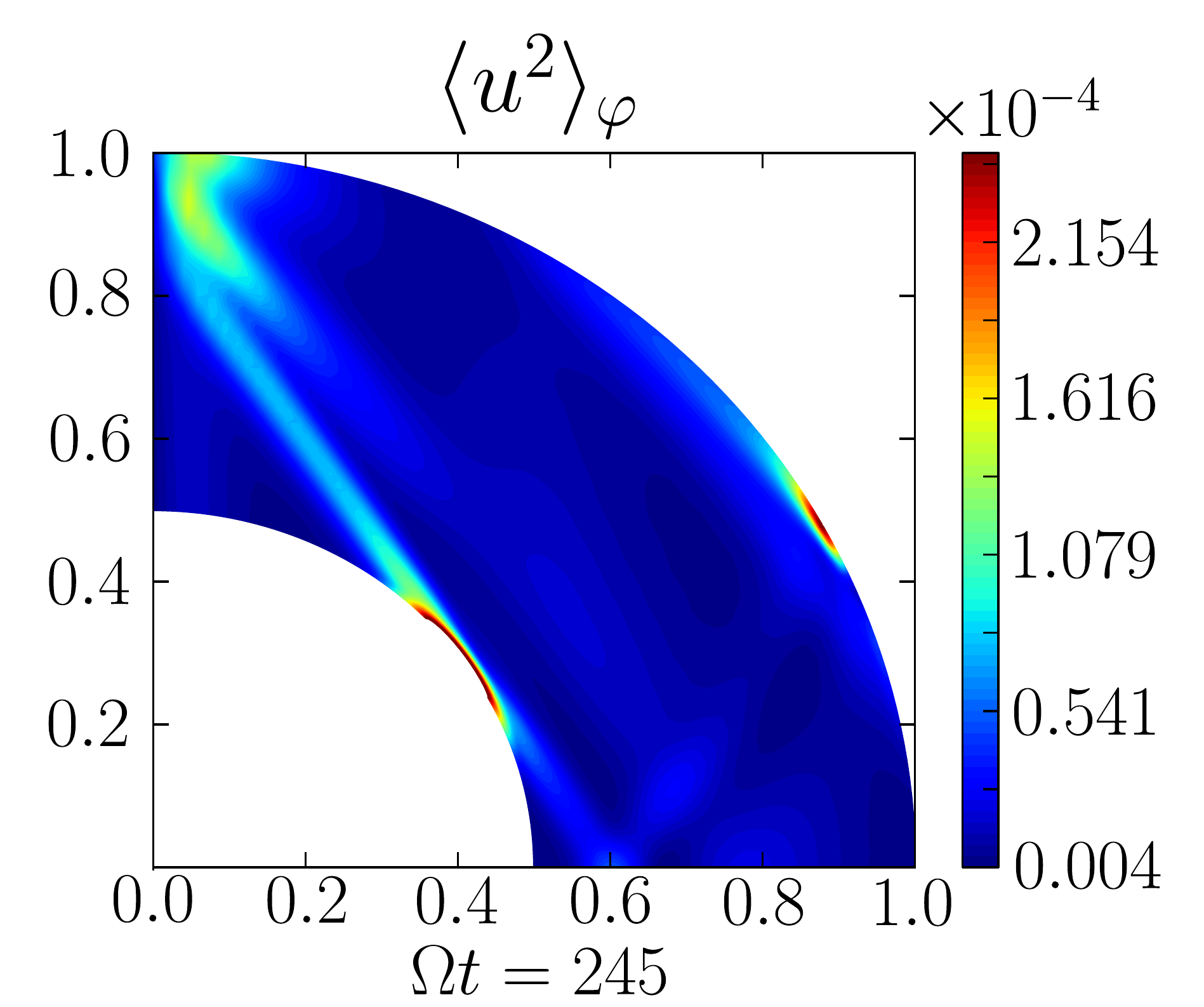}      
 \includegraphics[width=0.24\textwidth,clip]{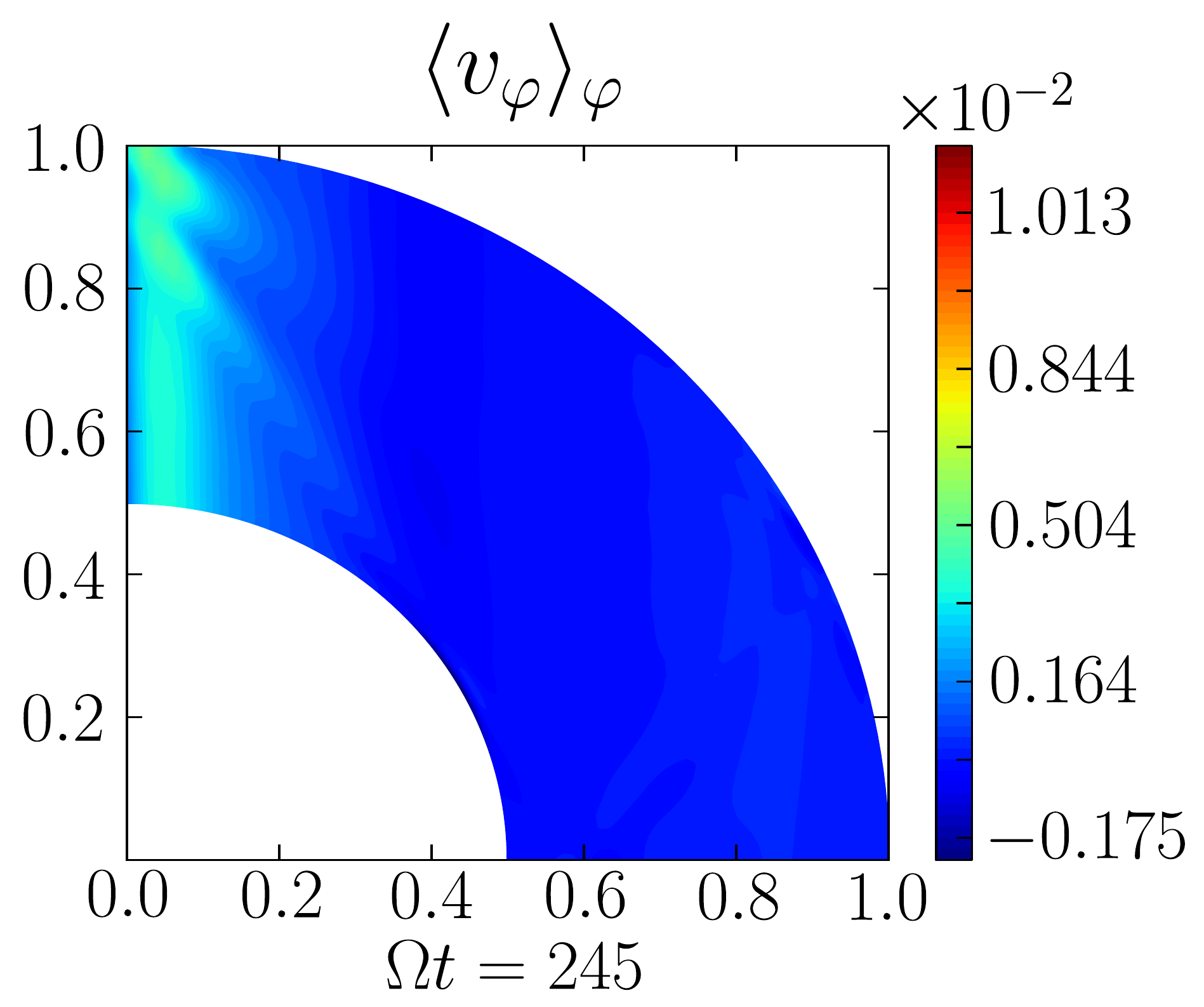}      
\includegraphics[width=0.24\textwidth,clip]{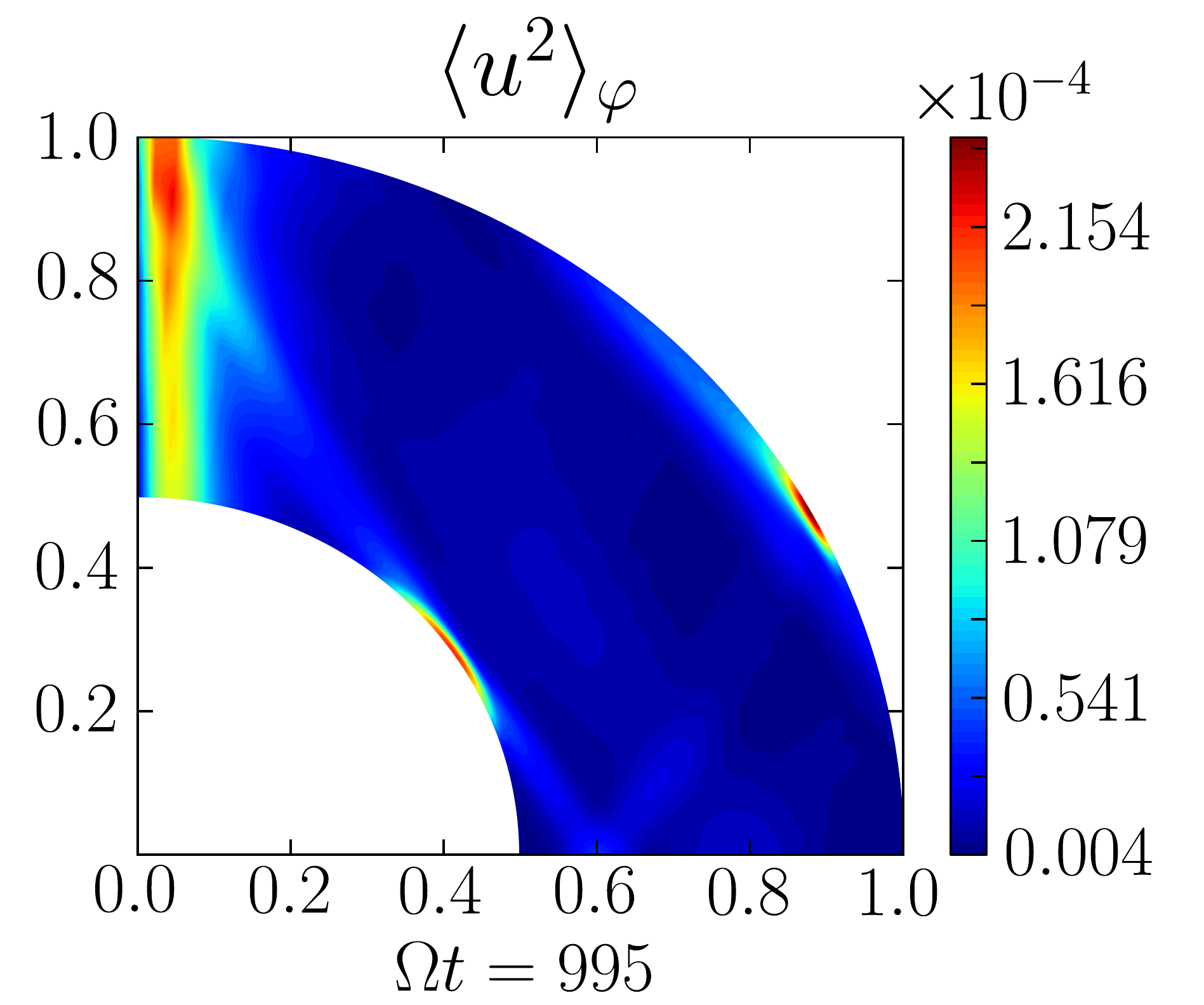}      
 \includegraphics[width=0.24\textwidth,clip]{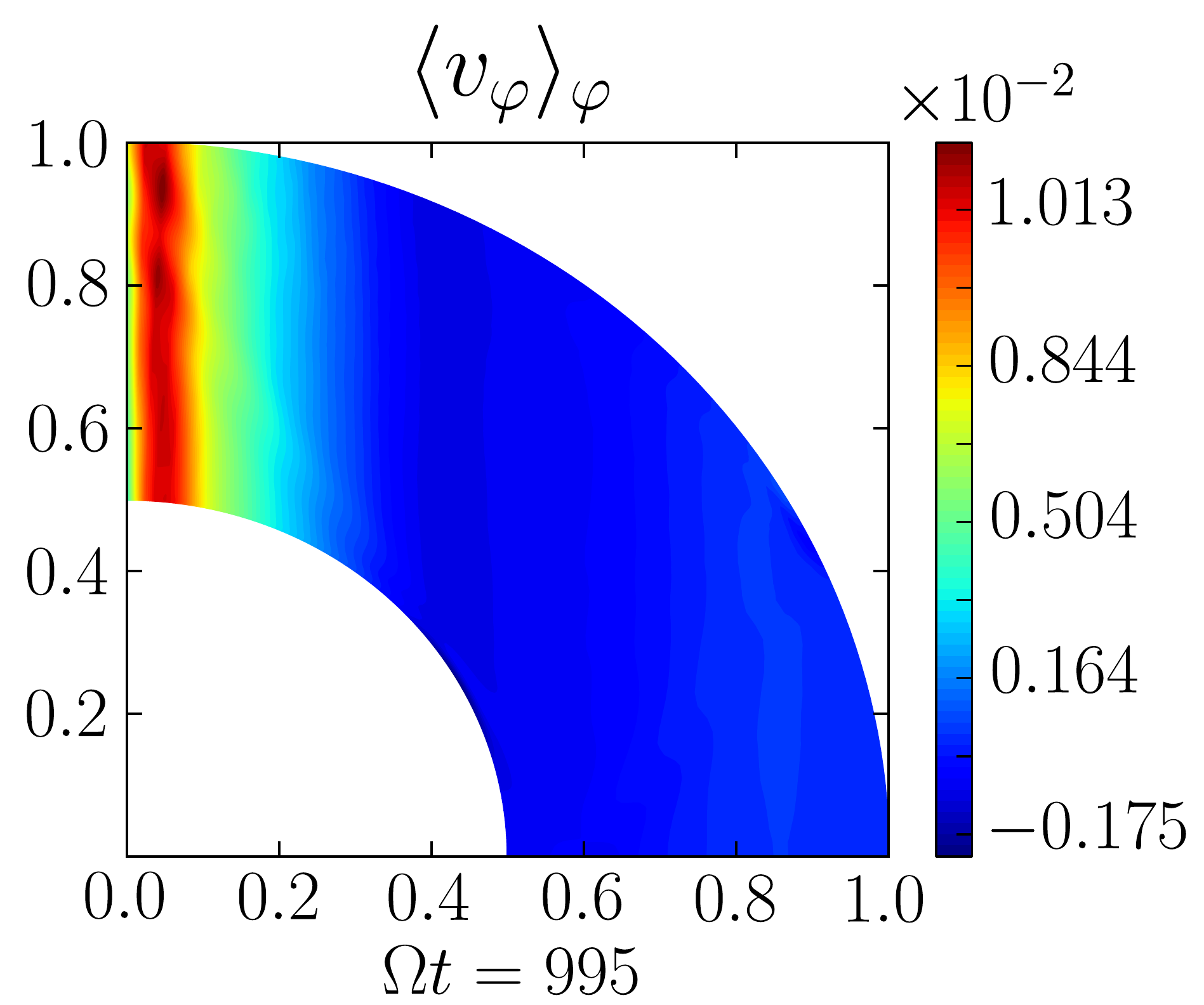}      
 \caption{Azimuthal average of the kinetic energy $u^2$ and azimuthal velocity $v_\varphi$ at times $\Omega t=245$ and $\Omega t=995$ in one quarter of the meridional plane (due to equatorial and axial symmetry) for $\omega=1.1$ and $C=0.009$. %MAKE BIGGER? ok
 }
  \label{astoul:fig2}
\end{figure}

We find that when all nonlinearities are included, the kinetic energy and tidal dissipation in our simulations are almost identical to those obtained in Paper~I (after appropriate rescaling). 
%when realising that $\sqrt{\frac{32\pi}{15}}A=\omega C$, with $A$ the tidal amplitude in Paper~I 
%\citep[after appropriate rescaling, see in particular][]{O2009,O2013}. 
In the middle panel of Fig.~\ref{astoul:fig1}, we show that the kinetic energy including all nonlinearities highly departs from the linear prediction for $\omega=1.05$ and $1.1$, with the departure being somewhat less pronounced when considering only the wavelike-wavelike nonlinearity. In both cases this departure from linear predictions may strongly depend on whether wave attractors are produced (i.e. wave focusing along limit cycles)
or to the presence of hidden large-scale flows %reminiscent of large-scale inertial modes in a full sphere %hidden beneath these localised waves beams
\citep[see][]{O2009, LO2021}. When one or both of these features are present the linear dissipation is enhanced, as we can see in the right panel of Fig.~\ref{astoul:fig1}. At the beginning of these %wavelike-wavelike
simulations, kinetic energy and dissipation converge towards their linear values, and then evolve away for $\omega=1.05$ and $1.1$ due to the establishment of strong zonal flows indicated by the energy in the differential rotation %where $E_\mathrm{dr}=    \frac{1}{2}\int_V\left[\langle u_\varphi\rangle_\varphi-\delta\Omega r \sin\theta \right]^2\mathrm{d}V$
\citep[colorbar, see Paper~1 and][for a definition]{T2007}. Since the total angular momentum is conserved here (as $\unw$ is perfectly maintained), a steady state is reached at the end of the simulation ($\max(\Omega t)=5000$).
The inhomogeneous deposition of kinetic energy (and angular momentum) near the rotation axis triggers cylindrical differential rotation from the pole to the equator as shown in Fig.~\ref{astoul:fig2} for the case with $\omega=1.1$. 
The final dissipation of wavelike-wavelike nonlinear simulations can be neatly recovered by running new linear simulations with a background cylindrical differential rotation, using the zonal flows extracted from the associated nonlinear simulation (right panel of Fig.~\ref{astoul:fig1} using the flow shown in e.g.~Fig.~\ref{astoul:fig2}). We have therefore demonstrated that the effects of zonal flows on tidal waves explains the departure of tidal dissipation % (tidal transfer) 
rates in these simulations from linear theoretical predictions assuming uniform rotation.

%%-------------------------------------
\section{Conclusions}
We have performed new nonlinear hydrodynamical simulations of tidally-forced inertial waves in an incompressible and adiabatic convective shell. Our approach is different from that in Paper~1 by the use of (a more realistic) effective tidal body force to excite inertial waves, along with stress-free and impenetrable boundary conditions. Within this framework, we analysed in detail the energy transfer terms between wavelike and non-wavelike tidal flows, and demonstrated that an “unphysical” energy flux through the spherical boundaries was responsible for the unrealistic angular momentum evolution obtained in certain cases in Paper~1. We have removed the mixed wavelike-non-wavelike nonlinearities responsible, which can be further justified by scaling arguments (since wave-wave nonlinearities are predicted to dominate for short wavelength inertial waves). Differential rotation in the form of zonal flows is triggered inside the shell due to wavelike-wavelike nonlinearities and there are important departures from linear predictions as observed in Paper~1, though the differences here are less pronounced in the cases we have presented. %, that may be explained by the conservation of the total angular momentum. 
Finally, we have also demonstrated that the departure of the nonlinear tidal dissipation from the linear prediction is explained by the zonal flows that are generated.

%we find that an energy transfer term between wavelike and non-wavelike flows in the energy balance and coming from mixed nonlinearities is 
 %causing “unphysical” non-wavelike flux through the boundaries, due to our adoption of spherical geometry. By switching off the relevant nonlinearities, we remove this artificial flux as well as unrealistic evolution of the angular momentum as seen in Paper~1. 
 
 %till, zonal flows are  triggered inside the shell due to the wavelike-wavelike nonlinearity and departure from linear prediction is observed as in Paper~1, though the difference is less pronounced, that may be explained by the conservation of the total angular momentum. Finally, the mitigation of nonlinear tidal dissipation compared to the linear one is shown to be driven by the development of sheared zonal flow.

% Optional acknowledgements
% -------------------------
\begin{acknowledgements}
This research has been supported by STFC grants ST/R00059X/1 and 
ST/S000275/1. Simulations were undertaken using the MagIC software on ARC4 and DiRAC, part of the High Performance Computing facilities at the University of Leeds and Leicester.
\end{acknowledgements}

%%-----------------------------
%%   Bibliography
%%-----------------------------
%%
%% The following lines are required when using BibTEX (strongly encouraged!):
\bibliographystyle{aa}  % A&A bibliography style file (aa.bst)
\bibliography{astoul_S6} % your references in file: Yourfile.bib

\end{document}